\documentclass{article}
\usepackage{spconf,amsmath,graphicx}
\usepackage{cite}
\usepackage{spconf,amsmath,graphicx}
\usepackage{amssymb}
\usepackage{mathrsfs}
\usepackage{amsmath}
\usepackage{comment}
\usepackage{here}
\usepackage{bm}
\usepackage{url}
\usepackage{booktabs}
\usepackage{enumitem}
\usepackage{siunitx}
\usepackage[multiple]{footmisc}

\title{Wearable SELD dataset: Dataset for sound event localization and detection using wearable devices around head}
\name{
Kento Nagatomo$^{\dagger}$,
Masahiro Yasuda$^{\ddagger}$,
Kohei Yatabe$^{\dagger}$,
Shoichiro Saito$^{\ddagger}$,
Yasuhiro Oikawa$^{\dagger}$
}

\address{$^\dagger$Waseda University, Tokyo, Japan \qquad\qquad
$^\ddagger$NTT Corporation, Tokyo, Japan
\vspace{-10pt}
}

\begin{document}
\ninept
\maketitle
\begin{abstract}
\vspace{-1pt}
Sound event localization and detection (SELD) is a combined task of identifying the sound event and its direction.
Deep neural networks (DNNs) are utilized to associate them with the sound signals observed by a microphone array.
Although ambisonic microphones are popular in the literature of SELD, they might limits the range of applications due to their predetermined geometry.
Some applications (including those for pedestrians that perform SELD while walking) require a wearable microphone array whose geometry can be designed to suit the task.
In this paper, for development of such a wearable SELD, we propose a dataset named \textit{Wearable SELD dataset}.
It consists of data recorded by 24 microphones placed on a head and torso simulators (HATS) with some accessories mimicking wearable devices (glasses, earphones, and headphones).
We also provide experimental results of SELD using the proposed dataset and SELDNet to investigate the effect of microphone configuration.
\vspace{-3pt}
\end{abstract}

\vspace{-2pt}
\section{Introduction}
\label{sec:intro}
%



\begin{figure}[t]
  \centering
  \vspace{-4pt}
  \centerline{\includegraphics[width=0.9\columnwidth,clip]{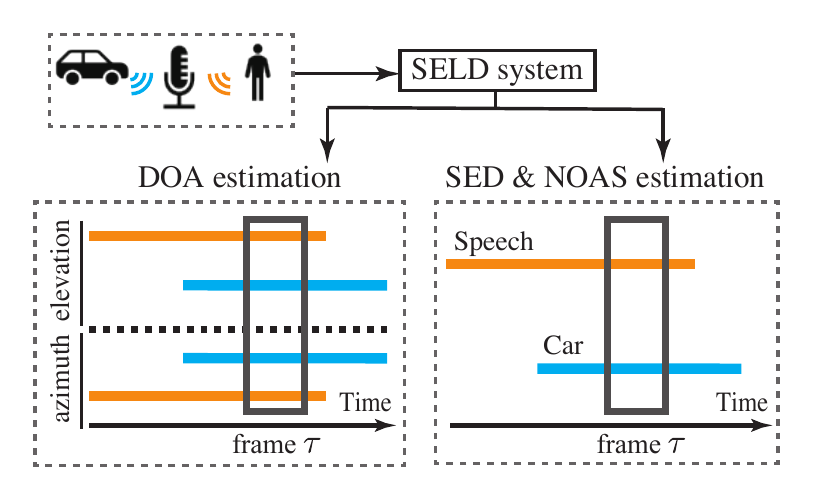}}
  \vspace{-14pt}
  \caption{Illustration of SELD (Sound Event Localization and Detection). DOA, NOAS, and SED stand for Direction-Of-Arrival, Number-Of-Active-Sources, and Sound Event Detection.}
  \label{fig:seld_overview}
\end{figure}

Detection and recognition of sound sources are essential for realizing intelligent systems, and thus a number of research has been devoted to them~\cite{detect_and_class, event_detect_real, scene_class, koizumi_sound_detec, SELD}.
Among many tasks, this paper focuses on sound event localization and detection (SELD), which is a recently proposed task that jointly addresses sound event localization (SEL) and sound event detection (SED)~\cite{SELD} (Fig.\:\ref{fig:seld_overview}). 
SEL identifies when and where a sound event occurred, which requires estimation of the number-of-active-sources (NOAS) and direction-of-arrival (DOA) from the observed sound.
SED identifies the sound event class and onset/offset times for each active source.
While SEL and SED had been addressed independently, recent studies have attempted to tackle them jointly \cite{SELD,sony_seld_2021, Parrish_2021, Nguyen_2021}.
Since SELD is a fundamental task for understanding the surrounding environment, it has a wide variety of applications including driver assistance~\cite{SmartCar1,SmartCar2} and security systems~\cite{drone}.
One important conceivable application is to help pedestrians to recognize the surroundings while walking.
With such application in mind, we consider a portable SELD system implemented by a wearable microphone array.

During the past years, SELD methods using deep neural networks~(DNNs) have been proposed~\cite{SELD,sony_seld_2021, Parrish_2021, Nguyen_2021}.
In particular, as in the other sound event/scene classification tasks~\cite{imoto}, DNN composed of convolutional network and recurrent network such as SELDNet (Fig.\:\ref{fig:seldnet}) has been used in the mainstream.
In most cases, the dataset in the first order ambisonics (FOA) format, which is generated by room impulse responses (RIRs) collected by spherical microphone arrays, has been used for the training~\cite{DCASE_dataset, Dcase2020dataset}.
Using FOA format is advantageous for extracting spatial information.
However, due to the predetermined array geometry, it restricts the variation of microphone arrangement required in a specific application.
Therefore, several other configurations have been considered in SEL to extend the range of applications \cite{Hearing_aid_beamformer, Three_three_hearing_aid,Helmet,smartglasses,Ad-hoc_microphone_array}.
Some datasets using wearable deivces are publicly available, but these are not suitable for SELD task because of the number of impulse responses (IRs) recordings.
Indeed, there are little research considering non-FOA configurations for SELD~\cite{MobilePhone, circular_array_seld}.

The configuration of the microphone array is an important factor for both the performance and usability of a SELD system.
A microphone array must be designed so that the SELD system is suitable for the target application.
For example, considering applications targeting pedestrians, small and wearable microphone array is preferable.
To develop a SELD system for such applications, a dataset recorded by the same microphone array is required.
However, to collect a set of data, the microphone setting must be fixed beforehand, which hinders from investigating the optimal microphone array configuration for SELD.
In particular, a SELD system using a wearable microphone array has not been proposed yet, and hence it is not easy to develop a wearable SELD system because the microphone configuration must be considered simultaneously.
To develop a wearable SELD system, a suitable dataset that contains a wide variety of microphone configurations is desired.

In this parper, we propose a new dataset named \textbf{Wearable SELD dataset}\footnote{\url{https://github.com/nttrd-mdlab/wearable-seld-dataset}}\footnote{\url{https://doi.org/10.5281/zenodo.6030111}} for development of wearable SELD.
We also realize SELD using it and evaluate the performance to give a recommendation of the microphone configuration.
That is, our contribution is twofold: (1) proposal of a new dataset, and (2) investigation of microphone arrangement using standard DNN architecture used in the field of SELD.
Our dataset is designed to cover some wearable devices around the head that is routinely used in our daily life, including glasses, earphones, and hearing aid.
In total, 24 microphones were placed on a head and torso simulators (HATS) with some accessories mimicking the devices.
Each microphone recorded individual data, and hence any combination of the 24 microphones can be used for the training of a SELD system.
To investigate the effect of microphone configuration, we performed experiments using the proposed dataset.
Results of the experiment performed for 240 combinations of microphones suggested that placing microphones behind the ears may degrade the performance of SELD.

\vspace{-2pt}
\section{Related works}
\vspace{-2pt}

\subsection{Sound event detection (SED)}
\vspace{-1pt}
In the field of SED, the mainstram methods are based on DNN, and hence a large dataset is required for training it.
Some popular datasets include AudioSet~\cite{AudioSet}, which is generated from YouTube data, and TUT Sound events~\cite{TUTsoundEvent2017}, which consists of binaural signals collected by in-ear earphones.
These datasets include information of event classes and timing but do not have information of spatial location and direction that are mandatory for SELD.

\vspace{-6pt}
\subsection{Sound event localization (SEL)}
\vspace{-1pt}
In the field of SEL, several microphone arrays have been used for developing systems applicable to various situations.
They are summarized in Table~\ref{tb:recording_devices}.
A distributed array~\cite{Ad-hoc_microphone_array} and a round table~\cite{Microsoft_roundtable} consist of fixed microphones and applied to SEL in the inner rooms.
A helmet-type microphone array~\cite{Helmet} was designed for industrial use cases, and hearing aid~\cite{Hearing_aid_beamformer, Three_three_hearing_aid} was designed for hearing impaired persons.
Some datasets collected by wearable microphone arrays are publicly opened.
One of the public datasets is Wearable Microphone Impulse Responses~\cite{wearable_device}, which consists of impulse responses from 160 microphones attached to the body and wearable accessories (hat, helmet, and headphone).
However, this dataset includes impulse responses captured at every $15^\circ$ azimuth angles for only one elevation angle. 
Since SELD has been tackled to estimate 3D direction for identifying the sound event location in detail, the variation of azimuth and elevation angles is not fine enough for SELD.

\vspace{-6pt}
\subsection{Sound event localization and detection (SELD)}
\vspace{-1pt}
SELD has been realized by DNNs trained with FOA format datasets \cite{Dcase2020dataset, DCASE2021dataset} that are collected by ambisonic microphones.
Since the FOA format can be converted to useful spatial quantities such as intensity vectors, recent SELD methods used it as an input feature and achieved high performance.
However, use of the FOA format might limit the range of applications because ambisonic microphones must be used in the SELD system.
Therefore, considering other type of microphone arrays should be valuable for widening the range of applications.
For example, a SELD system using handheld mobile phones~\cite{MobilePhone} has been developed for daily life situations.
One potentially important application of SELD is to assist a user while walking.
In such a case, a microphone array should be implemented as a wearable device.
Unfortunately, there has been no dataset using a wearale microphone array for SELD, but only publicly available ones are FOA format dataset \cite{Dcase2020dataset, DCASE2021dataset} and SECL-UMONS~\cite{circular_array_seld}.

\begin{figure}[t]
  \centering
  \centerline{\includegraphics[width=0.99\columnwidth,clip]{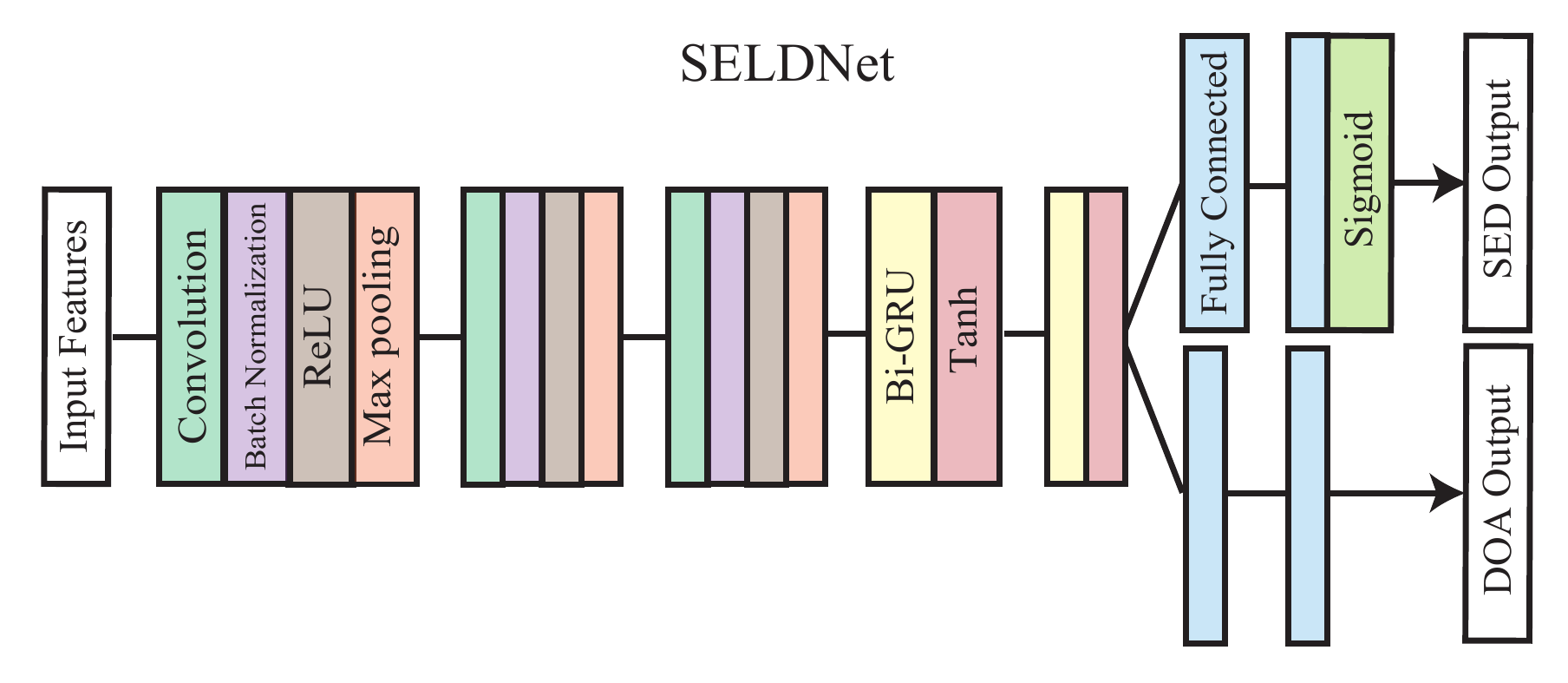}}
  \vspace{-17pt}
  \caption{Network architecture (SELDNet) used in the experiments.}
  \label{fig:seldnet}
  \vspace{-3pt}
\end{figure}

\begin{table}[t]
    \begin{center}
        \caption{Microphone arrays used in the literature. The bold fonts indicate that the datasets used in the references are publicly opened.}
        \vspace{-4pt}
        \label{tb:recording_devices}
        \begin{tabular}{@{}c|c@{}c@{}}
        \toprule
        & SEL & SELD \\ \midrule
        unwearable & \begin{tabular}{c}
        Distributed array\cite{Ad-hoc_microphone_array} \\ Round table\cite{Microsoft_roundtable}
        \end{tabular} & \begin{tabular}{c} \textbf{Spherical array}\cite{DCASE_dataset, Dcase2020dataset}\\ Mobile phone\cite{MobilePhone}\\ \textbf{Circular array}\cite{circular_array_seld}\end{tabular} \\ \midrule
        wearable & \begin{tabular}{c}Hearing aid\cite{Hearing_aid_beamformer,Three_three_hearing_aid} \\ Helmet\cite{Helmet}\\ Smart glasses\cite{smartglasses} \\ \textbf{Hat}\cite{wearable_device}\end{tabular} & \textbf{Proposed} \\ \bottomrule
        \end{tabular}
    \end{center}
    \vspace{-5pt}
\end{table}

\vspace{-2pt}
\section{Wearable SELD Dataset}
\vspace{-1pt}
To widen the application range of SELD, we openly provide a new dataset called \textit{Wearable SELD dataset}.
The dataset can be divided into two main parts.
One provides data collected by microphones around ears, which mimics the canal and ear hook type earphone.
The other part provides data for accessories around the head, mimicking glasses, headphones, and a neck speaker.
In addition, FOA format dataset is provided for comparison.

Specifically, Wearable SELD dataset consists of the following.
\begin{description}[leftmargin=15pt]
 \vspace{-2pt}
 \setlength{\parskip}{0cm} 
 \setlength{\itemsep}{1pt} 
 \item[Earphone type dataset:]
 This dataset contains data collected by 12 microphones placed around ears.
 Since earphones are one of the most familiar wearable devices in our daily life, 6 microphones were used for each device for detailed investigation.
 \item[Mounting type dataset:]
 This dataset contains data collected by 12 microphones placed around the head with some accessories mimicking glasses, headphone, and a neck speaker.
 The glasses and neck speaker can be worn together with earphones, and hence this dataset can be used with the earphone type dataset.
 \item[FOA format dataset:]
 This dataset was collected by ambisonic microphone to allow comparison with conventional methods using FOA format and those using the above datasets.
 \end{description}
These datasets except the FOA format dataset have three sub-datasets: anechioc version, reverberation version, and reverberation + noise version.
The FOA format dataset has only anechoic version.
In total, 7 sub-datasets are included in \textit{Wearable SELD dataset}.

\vspace{-6pt}
\subsection{Data collection}
\vspace{-1pt}
    The IRs were collected using a loudspeaker for emitting a chirp signal and omnidirectional microphones for its recording.
    We used HOSIDEN KUB4225 for the microphone. For the loudspeaker, we used a custom-made one whose model name is AMM Sound Lab CFX100F2506.
    The microphones were attached to the wearable accessories worn by a HATS.
    
\begin{figure}[t]
      \centering
      \centerline{\includegraphics[width=0.85\columnwidth,clip]{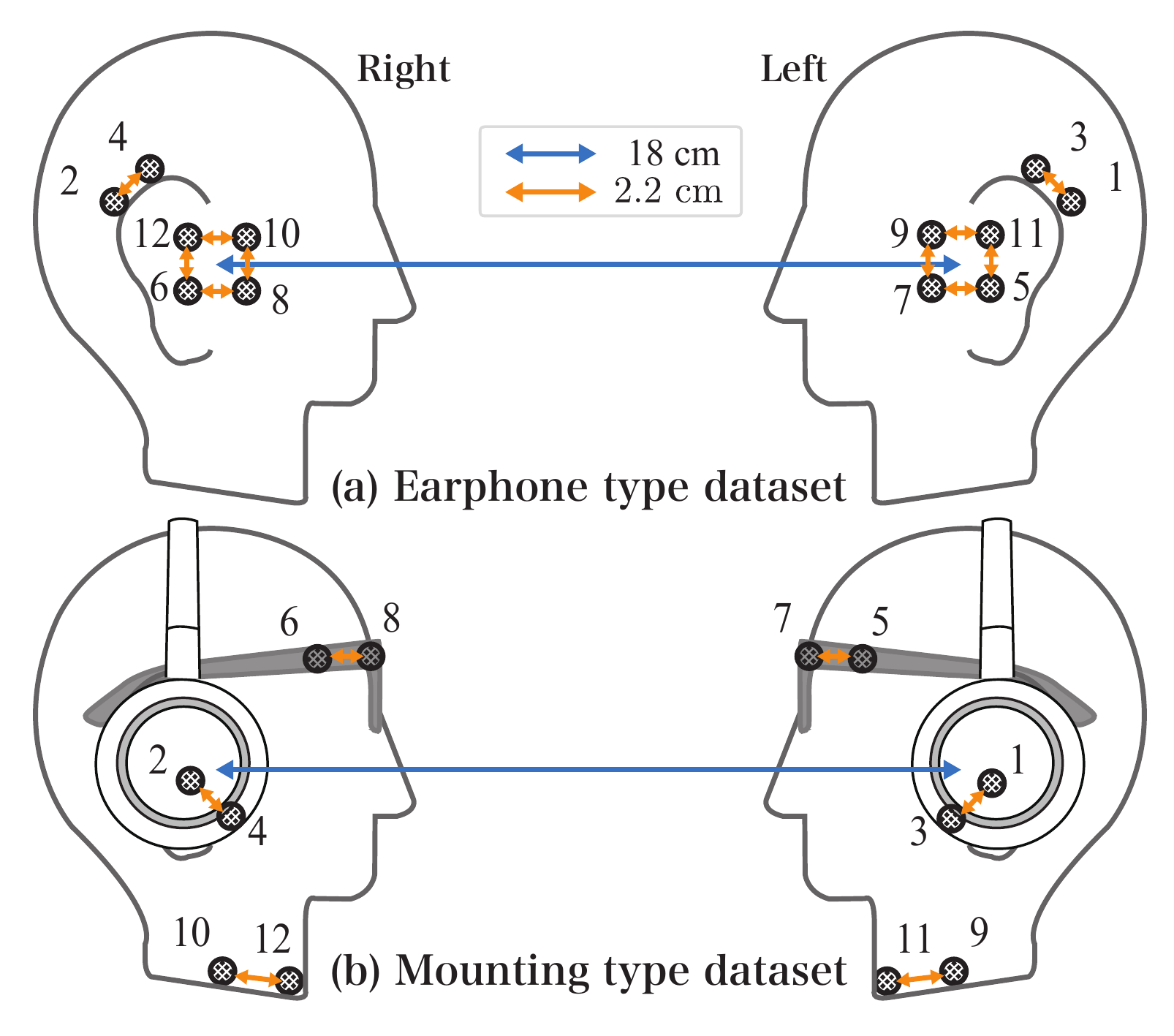}}
      \vspace{-13pt}
      \caption{Locations of the 12 microphones for the recording of each dataset. The number near each microphone indicates the channel ID.}
      \label{fig:mic_location}
      \vspace{-1pt}
    \end{figure}

    Our recording system handled 12 microphones simultaneously.
    Fig.\:\ref{fig:mic_location} (a) shows the microphone setting of the Earphone type dataset.
    A ear hook type earphone was covered by ch 1, 2, 3, 4; and a canal type earphone was covered by ch 5, 6, 7, 8, 9, 10, 11, 12.
    Fig.\:\ref{fig:mic_location} (b) shows the microphone setting of the Mounting type dataset.
    A headphone was covered by ch 1, 2, 3, 4; glasses were covered by ch 5, 6, 7, 8; and a neck speaker was covered by ch 9, 10, 11, 12.
    
    We collected IRs in a fully anechoic room and a variable reverberation room under two different reverberation times ($T_{60}$).
    The background noise of the fully anechoic room is \SI{-2}{\decibel}A and \SI{5}{\decibel}SPL at \SI{63}{\hertz}.
    The reverberation times of the variable reverberation room was adjusted to \SI{0.12}{\second} and \SI{0.41}{\second} at \SI{500}{\hertz}.
    Therefore, three room conditions were included in the Earphone type and Mounting type datasets.
    Note that the FOA format dataset was recorded only in the fully anechoic room.
    
\begin{figure}[t]
      \centering
      \centerline{\includegraphics[width=0.99\columnwidth,clip]{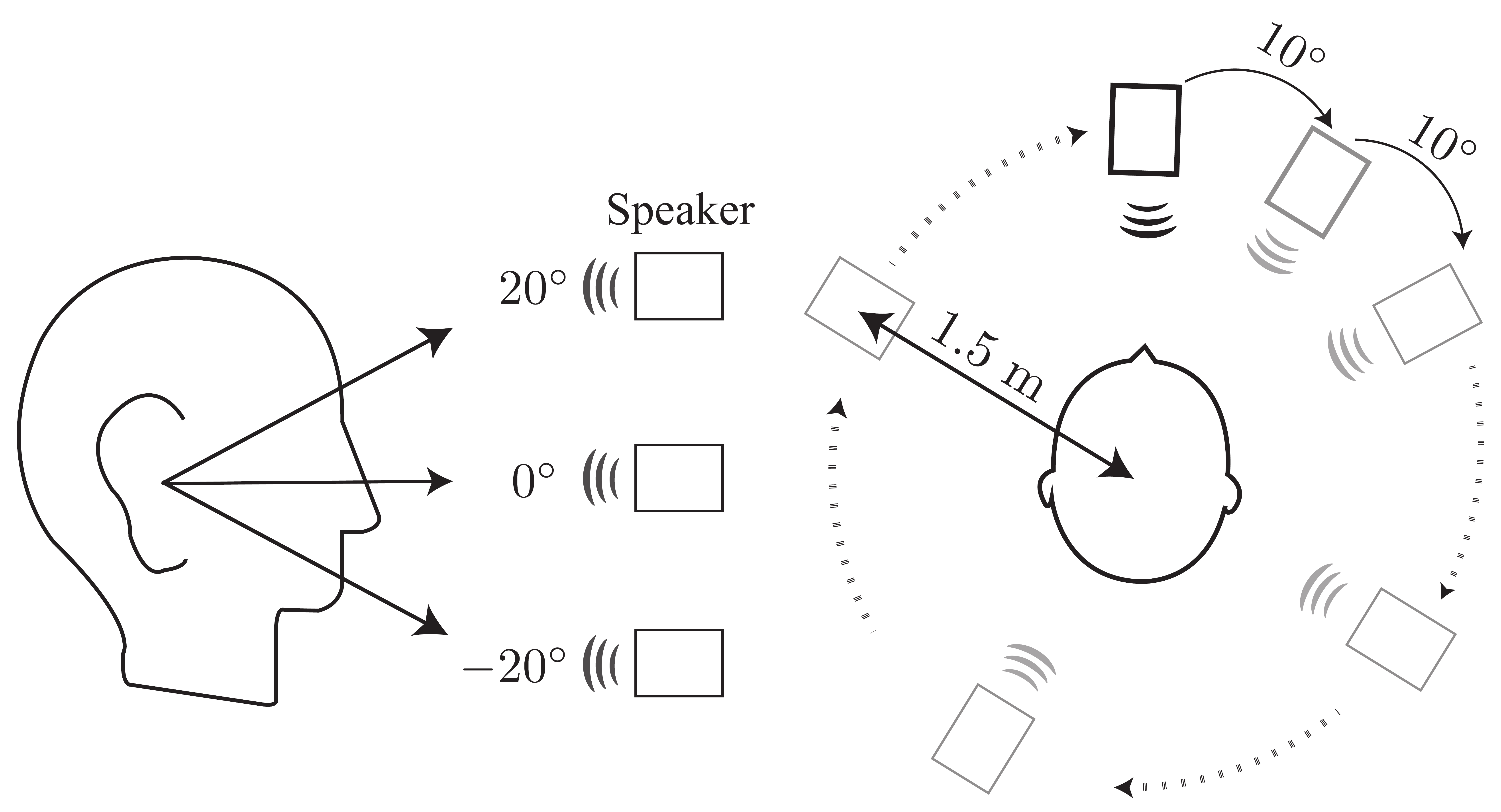}}
      \vspace{-8pt}
      \caption{Locations of the loudspeaker for data collection.}
      \label{fig:recording_env}
      \vspace{4pt}
    \end{figure}

    The IRs were recorded using the loudspeaker placed at 108 ($=3\times36$) discrete locations as shown in Fig.\:\ref{fig:recording_env}.
    The azimuth angle was discretized by \ang{10}, and hence 36 azimuth angles starting from \ang{0} were collected. The elevation angle was set to \ang{-20}, \ang{0}, \ang{20}.
    The distance between the loudspeaker and the center of the HATS was \SI{1.5}{\meter}.
    This setting was the same for all three room conditions.
    
    The reverberation-free recordings of sound events were recorded using the ominidirectional microphone in the fully anechoic room.
    The number of sound source classes is 12, and 20 samples were recorded for each class.
    The sound events consist of sounds generated by organ, piano, toy train, toy gun shot, metallophone, bicycle bell, security buzzer, shaker, handclap, woodblock, shaking bell, and hit drum.   The noise data was recorded in the reverberation room.
    White noise was emitted from 4 loudspeakers facing the walls and was recorded by each of the 12-channel microphone arrays in Fig.\:\ref{fig:mic_location}.

\vspace{-6pt}
    \subsection{Dataset synthesis}
    \vspace{-1pt}
    Wearable SELD dataset consists of 7 sub-datasets.
    Each of the Earphone type and Mounting type datasets contains three sub-datasets: anechoic version, reverberation version, and reverberation + noise version.
    The FOA format dataset contains only anechoic version.
    Each sub-dataset has the development set consisting of 400 data and the evaluation set consisting of 100 data.
    Each data is a one-minute long audio file sampled at \SI{48000}{\hertz}.
    
    A randomly selected reverberation-free recording of a sound event was convolved with a randomly selected IR.
    This spatialized sound event was temporally positioned at randomly generated time.
    Several spatialized sound events were added into an audio file, where the maximum number of overlapping events was 2 (half of the data in the development/evaluation sets has overlap).
    In an audio file, only one of the three room conditions was allowed.
    For the reverberated versions, each of the two reverberation conditions appears in half of the data in the development/evaluation sets (i.e., the development set is $200+200$, and the evaluation set is $50+50$).
    For the noisy version, the noise recorded with the corresponding room condition was added so that the average SNR became between 10 to \SI{20}{\decibel}.


\section{Experimental Investigation}

Using the proposed Wearable SELD dataset, we performed an experiment to investigate a suitable microphone setting for SELD.
From the 12 microphones in the Earphone type dataset, some combinations of 4 microphones were selected and evaluated. 
Out of ${}_{12}C_4 = 495$ combinations, 240 patterns of microphone settings were selected by the following rules.
First, each ear must have at least 1 microphone.
Second, essentially same microphone arrangements due to the symmetry should be removed for computational convenience.
These rules restrict the pattern of the number of microphones to 3vs1 and 2vs2. 
The resulted 240 patterns of settings were tested by training SELD systems.
In addition, those using the FOA format and the Mounting type datasets were tested as reference. 


\vspace{-6pt}
    \subsection{Experimental Setup}
    \vspace{-1pt}
    Our experiments were performed using SELDNet shown in Fig.\:\ref{fig:seldnet}.
    For the input features, amplitude spectrograms and phase diffrences (PD) were used.
    PD is defined as $\mathrm{PD}_{p,q\!} = \mathcal{W}(\angle {X}_{p} - \angle {X}_{q})$, where ${X}$ is a spectrogram, $\angle X$ is its phase, $p$ and $q$ represents the channel ID, and $\mathcal{W}$ denotes the phase wrapping operator.
    Since the data contains 4 signals, 4 amplitude spectrograms and 6 PD ($= {}_4C_2$) were concatenated and used as the input feature.
    The short-time Fourier transform (STFT) was implemented by the 2048-point Hann window with 960-point shifting step.
    The Adam optimizer~\cite{Adam} with the initial learning rate $\alpha=0.001$ was used for the training.
    The loss function and its parameter were the same as those in \cite{SELD}.
    
    For evaluation, DOA error (DE), frame recall (FR), error rate (ER), and F-score (F) were calculated because they were used in DCASE2019 Challenge Task3~(cf.~\cite{Metrics}).
    DE represents the error of the estimated angle, and FR represents the recall of NOAS estimation.
    These are the metrics related to SEL.
    On the other hand, ER and F are the metrics related to SED, where ER is the amount of error, and F is the harmonic average of accuracy and recall.

\vspace{-6pt}    
    \subsection{Results}
    \vspace{-1pt}

    \begin{table*}[t!]
    \vspace{-10pt}
    \begin{center}
        \caption{Performance of SELD using Earphone type dataset and FOA format dataset.} 
        \vspace{2pt}
        \begin{tabular}{l|c|c|cccc|cccc|cccc}
        \toprule
        & \multicolumn{2}{c}{Channels} & \multicolumn{4}{|c}{Anechoic} & \multicolumn{4}{|c|}{Reverberation} & \multicolumn{4}{c}{ Reverberation + noise} \\ \midrule
        & L & R &DE &FR  & ER & F  &DE &FR  & ER & F &DE &FR  & ER & F  \\ \midrule
        DE No. 1 &5,7,9 & 10 &5.61$^{\circ}$  &0.99  &0.017   &0.97& 6.90$^{\circ}$  &0.98  &0.025   &0.96 &11.1$^{\circ}$  &0.97  &0.045   &0.94 \\ 
        DE No. 2 &1,5,7 & 8 &5,72$^{\circ}$  &0.99  &0.018   &0.97 &7.57$^{\circ}$  &0.98  &0.029   &0.96 &12.0$^{\circ}$  &0.97  &0.048   &0.94  \\ \midrule
        ER No. 1 &1,7,9 & 8 &6.34$^{\circ}$  &0.99  &0.015   &0.97 &7.09$^{\circ}$  &0.98  &0.023   &0.97 &11.3$^{\circ}$  &0.97  &0.045   &0.94 \\ 
        ER No. 2  &7,9 & 8,12 &6.51$^{\circ}$  &0.99  &0.015   &0.97 &8.03$^{\circ}$  &0.98  &0.026   &0.97 &12.9$^{\circ}$  &0.97  &0.044   &0.94 \\ \midrule
        SELD No. 1 &5,7,11 & 6 &6.00$^{\circ}$  &0.99  &0.017   &0.98 &6.94$^{\circ}$  &0.98  &0.025   &0.97 &11.7$^{\circ}$  &0.97  &0.050   &0.94\\
        SELD No. 2 &5,7 &8,10&6.00$^{\circ}$  &0.99  &0.017   &0.98 &7.18$^{\circ}$  &0.98  &0.026   &0.97 &11.9$^{\circ}$  &0.97  &0.048   &0.94 \\ 
        SELD No. 240 &1,3 &2,4&6.77$^{\circ}$  &0.98  &0.026   &0.97 &8.18$^{\circ}$  &0.98  &0.028   &0.96 &14.6$^{\circ}$  &0.97  &0.054   &0.93\\ \midrule
        \multicolumn{1}{c|}{---} & \multicolumn{2}{c|}{FOA 4 ch} & 5.39$^{\circ}$  & 0.98  &0.021   &0.97 & ---  & --- & --- & --- & --- & --- & --- & --- \\ \bottomrule
        \end{tabular}
        \label{tb:seld_result}
        \vspace{-9pt}
    \end{center}
    
    \begin{center}
        \caption{Performance of SELD using Mounting type dataset.} 
        \vspace{2pt}
        \begin{tabular}{l|c|c|cccc|cccc|cccc}
        \toprule
        &\multicolumn{2}{c}{Channels} & \multicolumn{4}{|c}{Anechoic} & \multicolumn{4}{|c|}{Reverberation} & \multicolumn{4}{c}{ Reverberation + noise} \\ \midrule
        &L & R &DE &FR  & ER & F  &DE &FR  & ER & F &DE &FR  & ER & F  \\ \midrule
        Headphone &1,2 & 3,4 &6.41$^{\circ}$  &0.98  &0.019   &0.97& 7.23$^{\circ}$  &0.98  &0.023   &0.97 &10.2$^{\circ}$  &0.97  &0.049   &0.94 \\
        Glasses &5,6 & 7,8 &6.37$^{\circ}$  &0.99  &0.018   &0.98 &6.96$^{\circ}$  &0.98  &0.023   &0.97 &14.6$^{\circ}$  &0.97  &0.046   &0.94  \\
        Neck speaker &9,10 & 11,12 &6.49$^{\circ}$  &0.99  &0.018   &0.97 &6.85$^{\circ}$  &0.98  &0.022   &0.97 &11.8$^{\circ}$  &0.97  &0.049   &0.94 \\
        \bottomrule
        \end{tabular}
        \label{tb:seld_result_2}
        \vspace{-14pt}
    \end{center}
\end{table*}

Among the results for the 240 patterns of microphone setting, some extreme results are shown in Table~\ref{tb:seld_result}.
The best and second best models for the anechoic sub-dataset are selected for DE, ER, and SELD score.
The SELD score is obtained by the cumulative rank of the four metrics.
For the SELD score, we also show the worst result.

The result for the FOA format dataset is also shown as reference.
By comparing it with the best DE model, it can be seen that the wearable SELD model performed very close to that using the ordinary FOA format data (DE differs only \ang{0.22}).
Hence, an earphone-type wearable device has enough potential for SELD.

\begin{figure}[t!]
  \centering
  \includegraphics[width=0.89\columnwidth,clip]{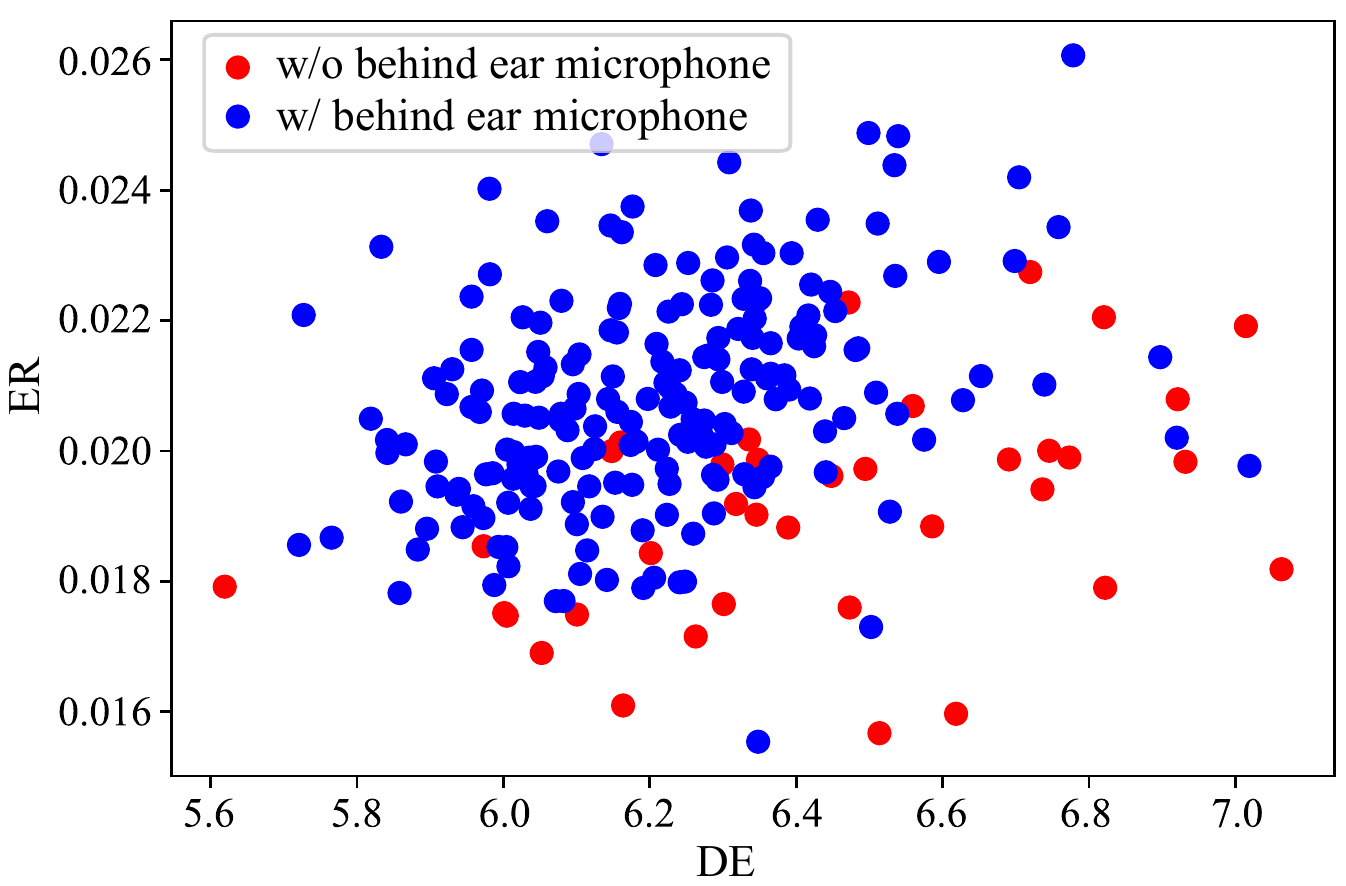}
  \vspace{-8pt}
  \caption{ER and DE of all 240 trained SELD models.}
  \label{fig:scatter_back_mic}
  \vspace{-2pt}
\end{figure}

By looking at the channel ID, it is interesting to note that the worst SELD model used ch 1, 2, 3, 4, which are the microphones placed behind the ears as shown in Fig.\:\ref{fig:mic_location}.
To analyze the data, all 240 results are summarized in Fig.\:\ref{fig:scatter_back_mic}, where the color indicates the use of a microphone behind an ear.
Since smaller DE and ER correspond to a better result, the left bottom is better in this figure.
From the figure, it can be seen that those using a microphone behind an ear tend to perform better for DE but worse for ER.

\begin{figure}[t!]
\vspace{-9pt}
  \centering
  \includegraphics[width=0.95\columnwidth,clip]{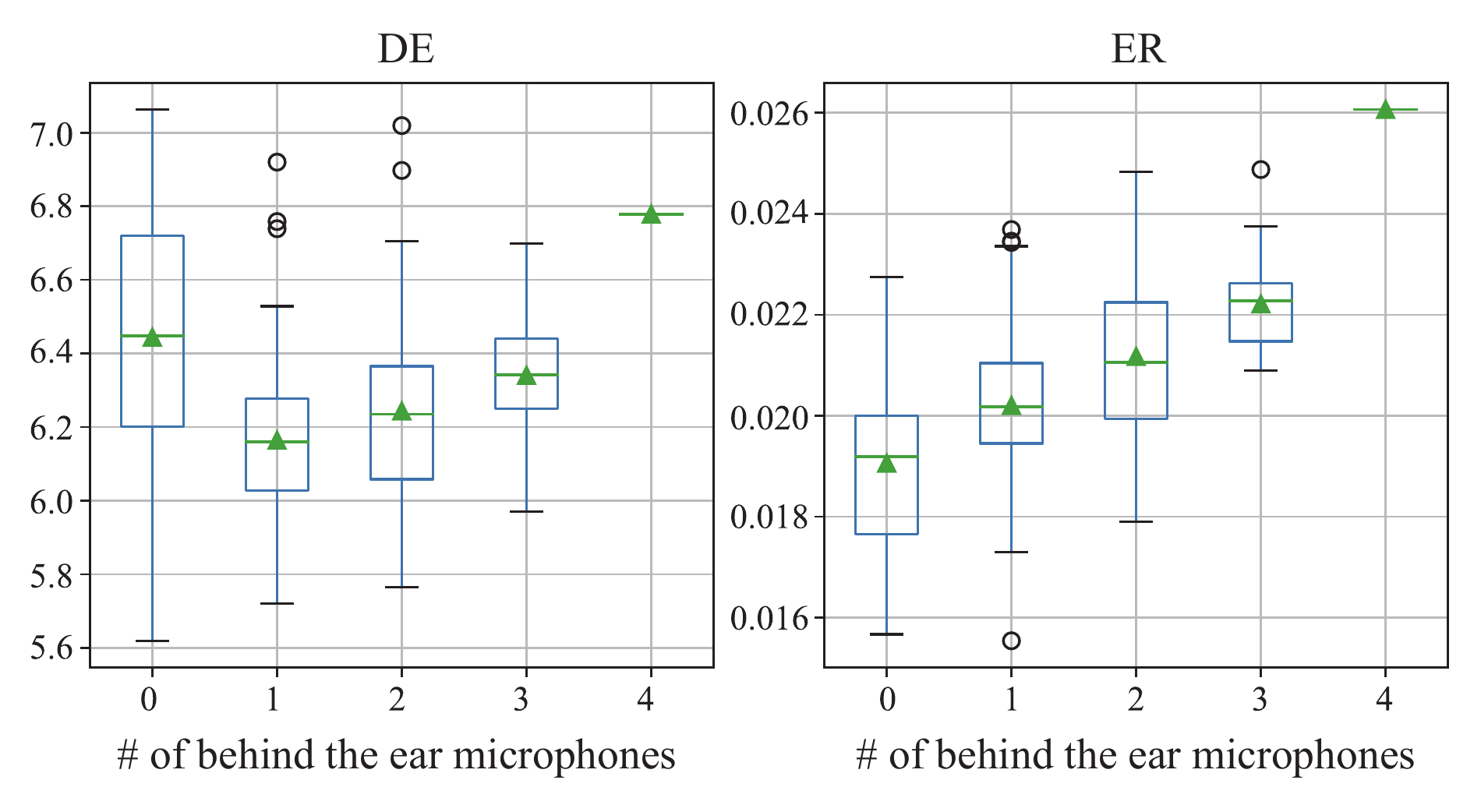}
  \vspace{-12pt}
  \caption{DE and ER grouped by number of microphones behind ear.}
  \label{fig:box_plot_back_mic}
  \vspace{-4pt}
\end{figure}

To further analyze the results focusing on the number of microphones behind the ears, we grouped the results according to it.
DE and ER are summarized in Fig.\:\ref{fig:box_plot_back_mic}.
From the right figure, we can see the clear relation between ER and the number of microphones behind the ears.
ER is better when no microphone is placed behind an ear, and ER becomes worse as the number of microphones behind the ears increases.
In contrast, in the left figure, such a simple relation cannot be observed for DE.
If we ignore the leftmost box, then the tendency similar to ER can be seen.
We can also see that the best DE model did not use a microphone behind an ear because the bottom end of the whisker grown from the leftmost box is lower than the others.
Therefore, for SELD using an earphone-type device, we recommend placing all microphones in front of the ears.

Our analysis was not able to reveal the reason why the leftmost box in the left figure of Fig.\:\ref{fig:box_plot_back_mic} has high deviation than the others.
Even so, we emphasize that the above analysis was made possible for the first time by the proposed Wearable SELD dataset.
Since this dataset is openly available, everyone can perform the similar experiments to investigate the performance of own SELD models.

SELD results for the Mounting type dataset are shown in Table~\ref{tb:seld_result_2}.
By comparing the results for the reverberation version with those with noise, we can see that the microphone setting using the glasses was more sensitive to noise (DE was more than doubled) than those using the headphone and neck speaker (DEs increased by a factor of 1.4 and 1.7, respectively).
Since DE \ang{10.2} is better than the others for the reverberant and noisy sub-dataset, a headphone seems to be a suitable device for a wearable SELD system in practice.

\vspace{-6.5pt}
\section{Conclusion}
\vspace{-4pt}
In this paper, we proposed a new dataset called Wearable SELD dataset.
We also performed an experiment and analysis using the proposed dataset to investigate the effect of microphone configuration.
The experimental results suggested that using microphones behind ears have risks of performance degradation for SELD.
Since the headphone model achieved the best DOA estimation under the noisy reverberant condition without degrading the other scores, a headphone seems to be a suitable wearable device for SELD in practice.

\vfill\newpage

\bibliographystyle{IEEEbib}
\bibliography{refs}

\begin{thebibliography}{10}

\bibitem{detect_and_class}
D.~{Stowell}, D.~{Giannoulis}, E.~{Benetos}, M.~{Lagrange}, and M.~D.
  {Plumbley},
\newblock ``Detection and classification of acoustic scenes and events,''
\newblock {\em IEEE Trans. Multimed.}, vol. 17, no. 10, pp. 1733--1746, 2015.

\bibitem{event_detect_real}
A.~{Mesaros}, T.~{Heittola}, A.~{Eronen}, and T.~{Virtanen},
\newblock ``Acoustic event detection in real life recordings,''
\newblock in {\em Proc. Eur. Signal Process. Conf. (EUSIPCO)}, 2010, pp.
  1267--1271.

\bibitem{scene_class}
D.~{Barchiesi}, D.~{Giannoulis}, D.~{Stowell}, and M.~D. {Plumbley},
\newblock ``Acoustic scene classification: Classifying environments from the
  sounds they produce,''
\newblock {\em IEEE Signal Process. Mag.}, vol. 32, no. 3, pp. 16--34, 2015.

\bibitem{koizumi_sound_detec}
Y.~{Koizumi}, S.~{Saito}, H.~{Uematsu}, Y.~{Kawachi}, and N.~{Harada},
\newblock ``Unsupervised detection of anomalous sound based on deep learning
  and the {N}eyman-{P}earson lemma,''
\newblock {\em IEEE/ACM Tran. Audio, Speech, and Lang. Process.}, vol. 27, no.
  1, pp. 212--224, 2019.

\bibitem{SELD}
S.~{Adavanne}, A.~{Politis}, J.~{Nikunen}, and T.~{Virtanen},
\newblock ``Sound event localization and detection of overlapping sources using
  convolutional recurrent neural networks,''
\newblock {\em IEEE J. sel. top. signal process.}, vol. 13, 2019.

\bibitem{sony_seld_2021}
K.~{Shimada}, N.~{Takahashi}, Y.~{Koyama}, S.~{Takahashi}, E.~{Tsunoo},
  M.~{Takahashi}, and Y.~{Mitsufuji},
\newblock ``Ensemble of accdoa- and einv2-based systems with d3nets and impulse
  response simulation for sound event localization and detection,''
\newblock Tech. {R}ep., DCASE2021 Challenge, 2021.

\bibitem{Parrish_2021}
P.~{Emmanuel}, N.~{Parrish}, and M.~{Horton},
\newblock ``Multi-scale network for sound event localization and detection,''
\newblock Tech. {R}ep., DCASE2021 Challenge, 2021.

\bibitem{Nguyen_2021}
T.~N.~T. {Nguyen}, K.~{Watcharasupat}, N.~K. Nguyen, D.~L. Jones, and W.~S.
  {Gan},
\newblock ``Dcase 2021 task 3: Spectrotemporally-aligned features for
  polyphonic sound event localization and detection,''
\newblock Tech. {R}ep., DCASE2021 Challenge, November 2021.

\bibitem{SmartCar1}
Y.~{Xu}, Q.~{Kong}, W.~{Wang}, and M.~D. {Plumbley},
\newblock ``Surrey-cvssp system for {DCASE}2017 challenge task4,''
\newblock in {\em Tech. Rep., DCASE2017}, 2017.

\bibitem{SmartCar2}
D.~{Lee}, S.~{Lee}, Y.~{Han}, and K.~{Lee},
\newblock ``Ensemble of convolutional neural networks for weakly-supervised
  sound event detection using multiple scale input,''
\newblock in {\em Tech. Rep., DCASE2017}, 2017.

\bibitem{drone}
X.~{Chang}, C.~{Yang}, X.~{Shi}, P.~{Li}, Z.~{Shi}, and J.~{Chen},
\newblock ``Feature extracted {DOA} estimation algorithm using acoustic array
  for drone surveillance,''
\newblock in {\em Proc. of IEEE 87th Veh. Technol. Conf.}, 2018.

\bibitem{imoto}
K.~{Imoto}, N.~{Tonami}, Y.~{Koizumi}, M.~{Yasuda}, R.~{Yamanishi}, and
  Y.~{Yamashita},
\newblock ``Sound event detection by multitask learning of sound events and
  scenes with soft scene labels,''
\newblock in {\em Proc. IEEE Int. Conf. Acoust. Speech Signal Process.
  (ICASSP)}, 2020.

\bibitem{DCASE_dataset}
S.~{Adavanne}, A.~{Politis}, and T.~{Virtanen},
\newblock ``A multi-room reverberant dataset for sound event localization and
  detection,''
\newblock in {\em Proc.of 4th Workshop on Detection and Classification of
  Acoustic Scenes and Events (DCASE)}, 2019.

\bibitem{Dcase2020dataset}
A.~{Politis}, S.~{Adavanne}, and T.~{Virtanen},
\newblock ``A dataset of reverberant spatial sound scenes with moving sources
  for sound event localization and detection,''
\newblock in {\em Proc. Workshop Detect. Classif. Acoust. Scenes Events
  (DCASE)}, November 2020.

\bibitem{Hearing_aid_beamformer}
M.~{Zohourian}, G.~{Enzner}, and R.~{Martin},
\newblock ``Binaural speaker localization integrated into an adaptive
  beamformer for hearing aids,''
\newblock {\em IEEE/ACM Trans. Audio Speech Lang. Process.}, vol. 26, no. 3,
  pp. 515--528, 2018.

\bibitem{Three_three_hearing_aid}
S.~{Goetze}, T.~{Rohdenburg}, V.~{Hohmann}, B.~{Kollmeier}, and K.~{Kammeyer},
\newblock ``Direction of arrival estimation based on the dual delay line
  approach for binaural hearing aid microphone arrays,''
\newblock {\em Int. Symp. Intell. Signal Process. Commun. Syst. ISPACS}, pp. 84
  -- 87, 2007.

\bibitem{Helmet}
P.~{Calamia}, S.~{Davis}, C.~{Smalt}, and C.~{Weston},
\newblock ``A conformal, helmet-mounted microphone array for auditory
  situational awareness and hearing protection,''
\newblock in {\em IEEE Workshop Appl. Signal Process. Audio Acoust. (WASPAA)},
  2017, pp. 96--100.

\bibitem{smartglasses}
D.~{Levin}, E.~{Habets}, and S.~{Gannot},
\newblock ``Near-field signal acquisition for smartglasses using two acoustic
  vector-sensors,''
\newblock {\em Speech Commun.}, vol. 83, 2016.

\bibitem{Ad-hoc_microphone_array}
W.~S. {Woods}, E.~{Hadad}, I.~{Merks}, B.~{Xu}, S.~{Gannot}, and T.~{Zhang},
\newblock ``A real-world recording database for ad hoc microphone arrays,''
\newblock in {\em IEEE Workshop Appl. Signal Process. Audio Acoust. (WASPAA)},
  2015, pp. 1--5.

\bibitem{MobilePhone}
P.~Pertila, Emre Çakir, Aapo Hakala, Eemi Fagerlund, Tuomas Virtanen,
  Archontis Politis, and A.~Eronen,
\newblock ``Mobile microphone array speech detection and localization in
  diverse everyday environments,''
\newblock in {\em arXiv:2106.14787}, 2021.

\bibitem{circular_array_seld}
M.~{Brousmiche}, J.~{Rouat}, and S.~{Dupont},
\newblock ``Secl-umons database for sound event classification and
  localization,''
\newblock in {\em IEEE Int. Conf. Acoust. Speech Signal Process. (ICASSP)},
  2020, pp. 756--760.

\bibitem{AudioSet}
J.~F. {Gemmeke}, D.~P.~W. {Ellis}, D.~{Freedman}, A.~{Jansen}, W.~{Lawrence},
  R.~C. {Moore}, M.~{Plakal}, and M.~{Ritter},
\newblock ``Audio set: An ontology and human-labeled dataset for audio
  events,''
\newblock in {\em IEEE Int. Conf. Acoust. Speech Signal Process. (ICASSP)},
  2017.

\bibitem{TUTsoundEvent2017}
M.~{Annamaria}, H.~{Toni}, and V.~{Tuomas},
\newblock ``{TUT} database for acoustic scene classification and sound event
  detection,''
\newblock in {\em 24th Eur. Signal Process. Conf. (EUSIPCO)}, 2016.

\bibitem{Microsoft_roundtable}
C.~{Zhang}, D.~{Florencio}, D.~E. {Ba}, and Z.~{Zhang},
\newblock ``Maximum likelihood sound source localization and beamforming for
  directional microphone arrays in distributed meetings,''
\newblock {\em IEEE Trans. Multimed.}, vol. 10, no. 3, pp. 538--548, 2008.

\bibitem{wearable_device}
R.~M. {Corey}, N.~{Tsuda}, and A.~C. {Singer},
\newblock ``Acoustic impulse responses for wearable audio devices,''
\newblock in {\em IEEE Int. Conf. Acoust. Speech Signal Process. (ICASSP)},
  2019, pp. 216--220.

\bibitem{DCASE2021dataset}
A.~{Politis}, S.~{Adavanne}, D.~{Krause}, A.~{Deleforge}, P.~{Srivastava}, and
  T.~{Virtanen},
\newblock ``A dataset of dynamic reverberant sound scenes with directional
  interferers for sound event localization and detection,''
\newblock {\em arXiv preprint arXiv:2106.06999}, 2021.

\bibitem{Adam}
D.~P. {Kingma} and J.~L. {Ba},
\newblock ``Adam: A method for stochastic optimization,''
\newblock in {\em Proc. IEEE Int. Conf. on Learn. Represent. (ICLR)}, 2015.

\bibitem{Metrics}
A.~{Mesaros}, S.~{Adavanne}, A.~{Politis}, T.~{Heittola}, and T.~{Virtanen},
\newblock ``Joint measurement of localization and detection of sound events,''
\newblock in {\em Proc. IEEE Workshop Appl. Signal Process. Audio Acoust.
  (WASPAA)}, 2019.

\end{thebibliography}

\end{document}